\documentclass[%
 reprint,
nofootinbib,
 amsmath,amssymb,
 aps,
prd,
floatfix,
]{revtex4-1}
\usepackage{graphicx,xcolor}
\usepackage{dcolumn}% Align table columns on decimal point
\usepackage{bm}% bold math
\usepackage{hyperref}% add hypertext capabilities
\usepackage{url}
\begin{document}

\preprint{Imperial-XXXX}

\title{Improved limits on a stochastic gravitational-wave background and its anisotropies from Advanced LIGO  O1 and O2 runs}% 

\author{A.~I.~Renzini}
\email{arianna.renzini15@imperial.ac.uk}
\author{C.~R.~Contaldi}

\affiliation{%
 Theoretical Physics, Blackett Laboratory, Imperial College, London, SW7 2AZ, United Kingdom.
}%


\date{\today}% It is always \today, today,

\begin{abstract}
We integrate the entire, publicly available, Advanced LIGO (ALIGO) data set to obtain maximum-likelihood constraint maps of the Stochastic Gravitational-Wave Background (SGWB). From these we derive limits on the energy density of the stochastic background $\Omega_{\rm GW}$, and its anisotropy. We find 95\% confident limits $\Omega_{\rm GW} < 5.2\times 10^{-8}$ at $50$ Hz for a spectral index $\alpha=2/3$ consistent with a stochastic background due to inspiral events and $\Omega_{\rm GW} < 3.2\times 10^{-7}$ for a scale (frequency) invariant spectrum. We also report upper limits on the angular power spectra $C_\ell$ for three broadband integrations of the data. Finally we present an estimate where we integrate the data into ten separate spectral bins as a first attempt to carry out a model-independent estimate the SGWB and its anisotropies.
\end{abstract}

\maketitle

%\tableofcontents

\section{Introduction} \label{sec:Intro}

The ability of the Laser Interferometer Gravitational-wave Observatory (LIGO) and Virgo Collaboration detectors in measuring the signal emitted during the merger of massive objects is beyond doubt with new events routinely announced during data-taking runs \cite{Abbott2016a,Abbott2017,Abbott2017b,Abbott2017a,Abbott2017c}. These measurements make use of the coherent detector's ability to measure the phase information in the signal of individual events. These events represent the strongest signals, within the relevant frequency and polarisation range, that are reaching the observatory. By definition these are necessarily the outlying events in the distribution of signals from such mergers and one would expect that there should also be a background of much lower signal-to-noise events that are more frequent. This background would also include the much more stationary signal emitted during the inspiraling phase of binary systems. For a sufficiently dense temporal distribution of events of sufficient duration the signal will add incoherently to form a stochastic background where the phase information in the observed time-stream becomes random and therefore redundant.

The expected spectral dependence of such a stochastic gravitational-wave background (SGWB) (see e.g. \cite{Christensen2018}) may not make it an obvious target for LIGO observations given their relatively high frequency but it is nonetheless interesting to develop methods that integrate the data efficiently assuming an incoherent signal. There may, for example, be an unexpected source for such a background but at the very least it is important to place upper limits at the frequencies LIGO is currently sensitive to. Additionally, future observatories sensitive to much lower frequencies such as the Laser Interferometer Space Antenna (LISA)~\cite{Amaro2017}, and the Einstein Telescope (ET)~\cite{Sathyaprakash2011}, are virtually guaranteed to detect a stochastic background of galactic and perhaps extra-galactic astrophysical sources~\cite{Regimbau2011}. There is also the potential for detecting stochastic backgrounds of much earlier sources such as early phase transitions or a primordial background generated during a period of inflation~\cite{Caprini2018,Battye1997}.

The background closest to present sensitivity levels for ground-based detectors~\cite{Moore} is the astrophysical background dominated by inspiral signals due to the mergers of massive objects such as black holes and neutron stars. Although there is still considerable uncertainty in the level of this background, there is a real possibility that the next upgrade of ground based detectors will reach the sensitivity required for a detection~\cite{Christensen2018}.

In the case of LISA there is the certainty that a stochastic background due to the persistent, incoherently superposed, residual signals from galactic binaries will be present. To separate this signal from others of interest the first step will be to use any distinct spectral dependence but at the same time it will be effective to make use of any angular resolution in the observations to separate out the signal on the sky. In order to do this methods that resolve both frequency and directionality of the observations must be developed. We have introduced one particular method in \cite{Renzini2018} and shown how it can be used to obtain constraints using LIGO observations \cite{Renzini2019} whilst similar efforts using different techniques have been described in the literature \cite{LIGOS4,LIGOS5,LIGO2016a,TheLIGOScientificCollaboration2019mono,TheLIGOScientificCollaboration2019}. In fact, the anisotropy itself may contain information that could be cross-correlated with large scale structure catalogues to determine its origin \cite{Contaldi:2016koz,Cusin2019}. 

In this paper we report on the limits obtained from the application of our map-making algorithm \cite{Renzini2018} to the combined integration of the the first (O1) and second (O2) Advanced LIGO runs~\cite{Renzini2019}. We present two independent limits on the SGWB: broadband limits obtained by integrating sky-projected data over the entire frequency range, and a novel model-independent method of extracting spectral limits where the integration is carried out over distinct frequency bins in order to obtain an optimal spectral decomposition of the data. We briefly review our method in Sec.~\ref{sec:method} whilst referring the reader to \cite{Renzini2018} for more details. Sec.~\ref{sec:data} describes the application to the LIGO data. We summarise some characteristics of the noise in Sec.~\ref{sec:noise} and discuss broadband results in Sec.~\ref{broad} and the model-independent spectral decomposition in Sec.~\ref{sec:segm10}. In both cases we distinguish between the monopole of the background as encoded by the energy density in gravitational waves normalised to the critical density $\Omega_{\rm GW}$ {\sl and} the anisotropies about the average background. We conclude with a discussion of our findings in Sec.~\ref{sec:disc}.

\begin{figure*}[t]
\begin{minipage}{0.5\textwidth}
  \centering
  \includegraphics[width=0.98\linewidth]{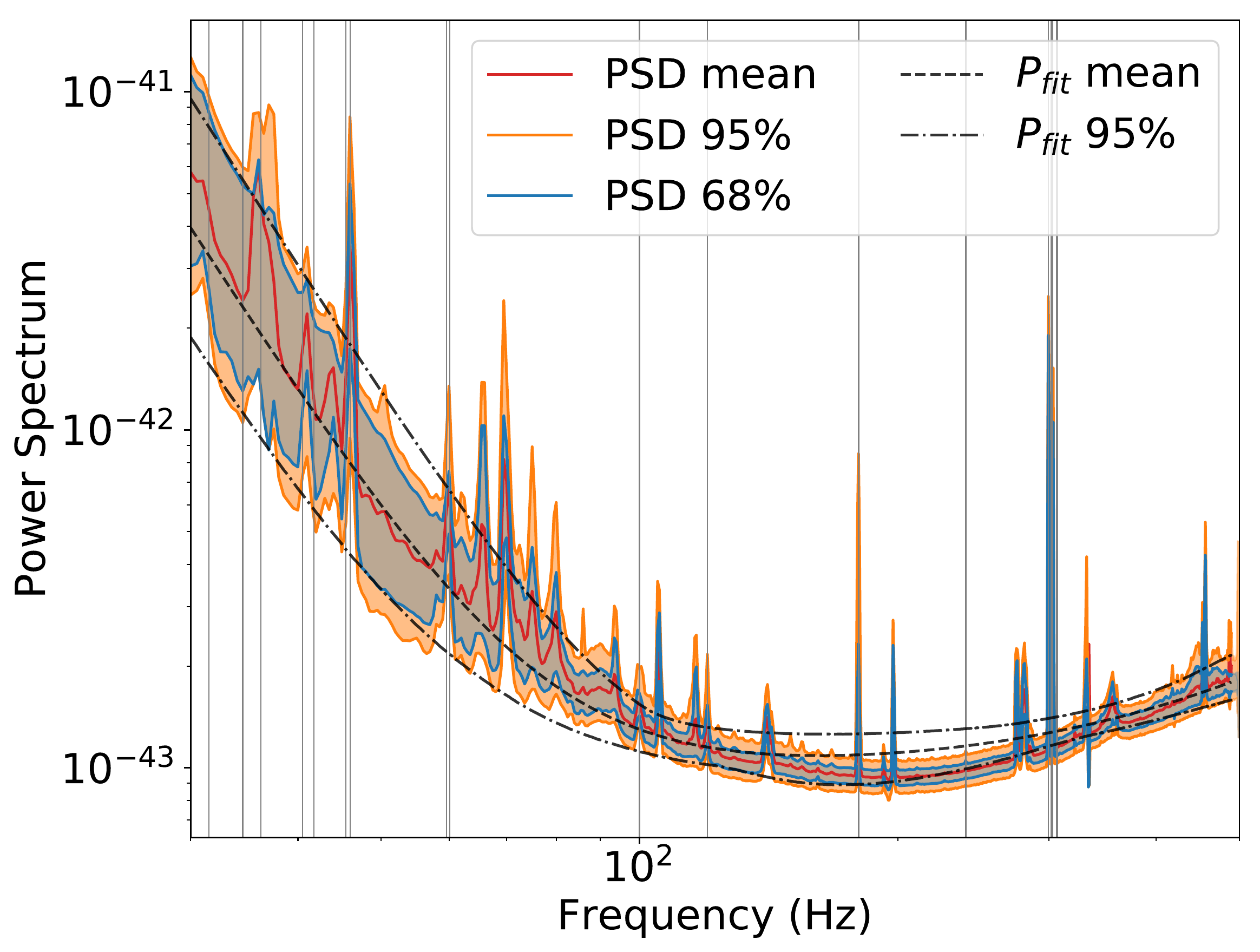}
\end{minipage}%
\begin{minipage}{0.5\textwidth}
  \centering
  \includegraphics[width=0.98\linewidth]{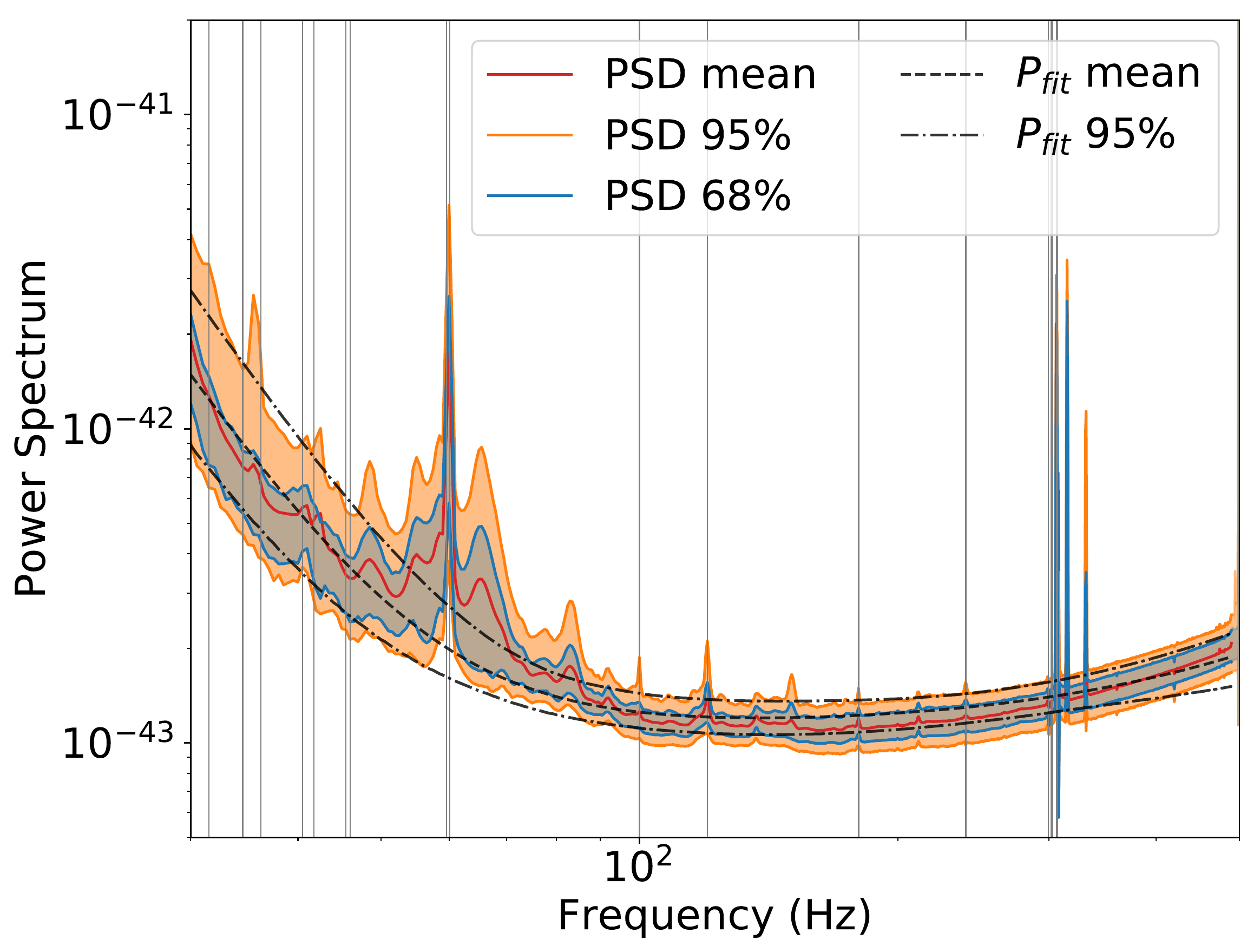}
\end{minipage}
\caption{Distribution of noise spectra (PSD) for Hanford ({\sl left}) and Livingston ({\sl right}) detectors. The shaded areas represent the 95\% and 68\% contours of the scatter in 60 second segments. The dashed lines show the mean and 95\% contours for the scatter in the three parameter fit to the noise in each 60 second segment. Known systematic harmonics are shown in the vertical lines. Other noise lines are unidentified and shift frequency over time producing wider ``forests'' in specific regions. The noise fitting procedure masks a number of these regions in order to reduce the bias in the parametrised model.}\label{fig:epsart}
\end{figure*}

\section{Method}\label{sec:method} 

In this paper we implement the mapping procedure described in detail in \cite{Renzini2018}, previously applied to Advanced LIGO open data set O1 in \cite{Renzini2019}, to the second Advanced LIGO data set, O2, which spans from 30 November, 2016 to August, 2017. The algorithm extracts the maximum-likelihood estimate for the SGWB integrating over filtered time-stream data cross-correlated between a pair of detectors.
Our pipeline identifies time-coincident LIGO Hanford and Livingston data blocks which pass a combination of quality flags, similarly to~\cite{Renzini2019}. This data cut reduces the viable O2 time-stream to 117.6 days which, when added to the coincident O1 data, amounts to a total of 166.2 days. The blocks are then divided into 60 second segments labelled $\tau$, tapered using a narrow cosine window of width 3 seconds to reduce edge effects and Fourier transformed to the frequency domain. Pre-estimated problematic frequencies and known harmonics are notch filtered out of the segments \cite{Covas2018} which are then band passed in the frequency range $[30.0,\,480.0]$ Hz. We then fit a three parameter analytic curve to the power spectrum density of each segment to obtain a specific weight function, as in~\cite{Renzini2019}, and discard segments which are not consistent with the model. This procedure cuts a further 15\% of the remaining data. For details about the notching procedure and the fitting results see Sec.~\ref{sec:data}.

The maximum-likelihood map of the SGWB intensity is then constructed following  \cite{Renzini2018}. The algorithm accumulates a noise-weighted projection of the time-stream data into map $z_p$  and  weight matrix $M_{pp'}$ over time,
\begin{equation}\label{method}
z_p = \sum_{\tau, \,f} A^{\tau \star}_{pf}\, N_f^{-1}\,d_f^\tau\,,
\end{equation}
\begin{equation}\label{method2}
 M_{pp'} = \sum_{\tau, \,f} A^{\tau \star}_{pf}\, N_f^{-1} \, A^\tau_{fp'}\,,
\end{equation}
where $p$ denotes the pixel number on the discretised sky following the {\tt HealPix} convention\footnote{\url{https://healpix.jpl.nasa.gov/}.}~\cite{Healpix}, $d^\tau_f$ is the frequency domain cross-correlation data during time segment $\tau$, $A^\tau_{pf}$ is the time-dependent operator that projects the data from the frequency domain onto sky pixels, and $N_f$ is the noise covariance of the data, assumed to be diagonal in frequency. We work at a {\tt HEALPix} resolution $N_{\rm side}=8$ corresponding to $n_{\rm pix}=768$ pixels on the sky. This corresponds to a pixelisation scale $\sim 7$ degrees or a Nyquist scale at multipole $\ell\sim 32$. For the purpose of visualising our results we over-smooth the maps using a Gaussian beam of 10 degrees full width half maximum (FWHM) in order to suppress noise at the resolution scale.

We have also assumed a discretisation in frequency indexed by $f$, working at the standard LIGO down sampled rate of 4096 Hz. The operator $A^\tau_{pf}$ contains all the details of the directional sensitivity of the detector combination on the sky as a function of time and also weights the data with an assumed frequency dependence $E(f)$ for the intensity on the sky. This is equivalent to assuming that the directional and frequency dependence of the SGWB intensity separate as
\begin{equation}
    {\cal I}(f,\hat {\mathbf n}_p) = E(f)\,I(\hat {\mathbf n}_p)\,.
\end{equation}

The map $z_p$ is a projection to inverse noise units. To obtain a correctly normalised, maximum-likelihood map one needs to weight this projection by the pixel domain noise covariance matrix $M^{-1}_{pp'}$.
The maximum-likelihood estimate of the sky intensity $I$ is then given by
\begin{equation}\label{eq:solution}
{\hat I}_p = \sum_{p'} M^{-1}_{pp'}\,z_{p'}\,.
\end{equation}
In our convention, pixel indices refer to fixed directions in galactic coordinates. This assumes that the directional sensitivity included in the operator $A^\tau_{pf}$ is rotated from celestial to galactic coordinates accordingly. A detailed discussion of how the operator is constructed is given in~\cite{Renzini2018}.

Since the signal is expected to be entirely sub-dominant in the frequency range being analysed here and given the duration of the 60 second time segments, the power spectrum of the data time-stream itself can be considered an unbiased estimate of the frequency domain noise covariance $N_f$.

In contrast to similar efforts by the LIGO collaboration \cite{Abadie2011,LIGO2016a,TheLIGOScientificCollaboration2019} our method is based on a pixel domain projection instead of a direct projection into the spherical harmonic domain. This choice is driven by the conditioning of the inversion employed in (\ref{eq:solution}). At finite resolution, we find the conditioning is better with this choice due to the complicated mixing of spherical harmonic domain modes implied by the observation of the sky using detectors with non-compact beams \cite{Renzini2018}. Once obtained, the maximum-likelihood maps can be decomposed into spherical harmonics in order to compare with results in \cite{TheLIGOScientificCollaboration2019} and we do this below although the comparison can only be approximate given the very different ways in which the resolution cutoff is applied when projecting directly to different domains. 
We also choose to work in a general sky frame, rather than fixing a specific detector-based frame. Choosing the detector frame may lead to substantial simplifications, see for example~\cite{Ain2015}, however working in sky coordinates allows us to set up a general framework for a network of more than two detectors, as described in~\cite{Renzini2018}. The independence and complementarity of this approach with respect to the LIGO one will be very useful as we approach the sensitivity levels required for a detection.
 %example the merger detections reported in~\cite{Venumadhav2019}.}

The choice of spectral dependence $E(f)$ for the intensity on the sky results in strong model dependence of the maximum-likelihood maps. The standard, so far, has been to present results as a function of spectral index $\alpha$ defining a power law assumed shape $E(f)\sim (f/f_0)^{\alpha-3}$ where $f_0$ is the specific reference frequency at which the estimate is made. In our model-dependent analysis described in Sec.~\ref{broad} we consider three theoretically motivated shapes: $\alpha = 0$, which corresponds to a flat frequency dependence in $\Omega_{\rm GW}$ and thus a template for a background of cosmological origin~\cite{Caprini2018}; $\alpha = 2/3$, which is the expected dependence for an inspiral dominated background~\cite{Sesana2008}; and $\alpha = 3$ which corresponds to the optimal spectral weight given the shape of the data noise. In Sec.~\ref{sec:segm10} we report constraints, for the first time, of a model independent spectral analysis where we break up the signal into narrow band segments. We apply the map-making procedure in each band, obtaining a set of independent SNR and noise maps for the intensity of gravitational-wave signal; we then use these to obtain an {\sl unbiased} estimate for the spectral shape of $\Omega_{\rm GW}$.   

\begin{figure}[t]
\includegraphics[width = \linewidth]{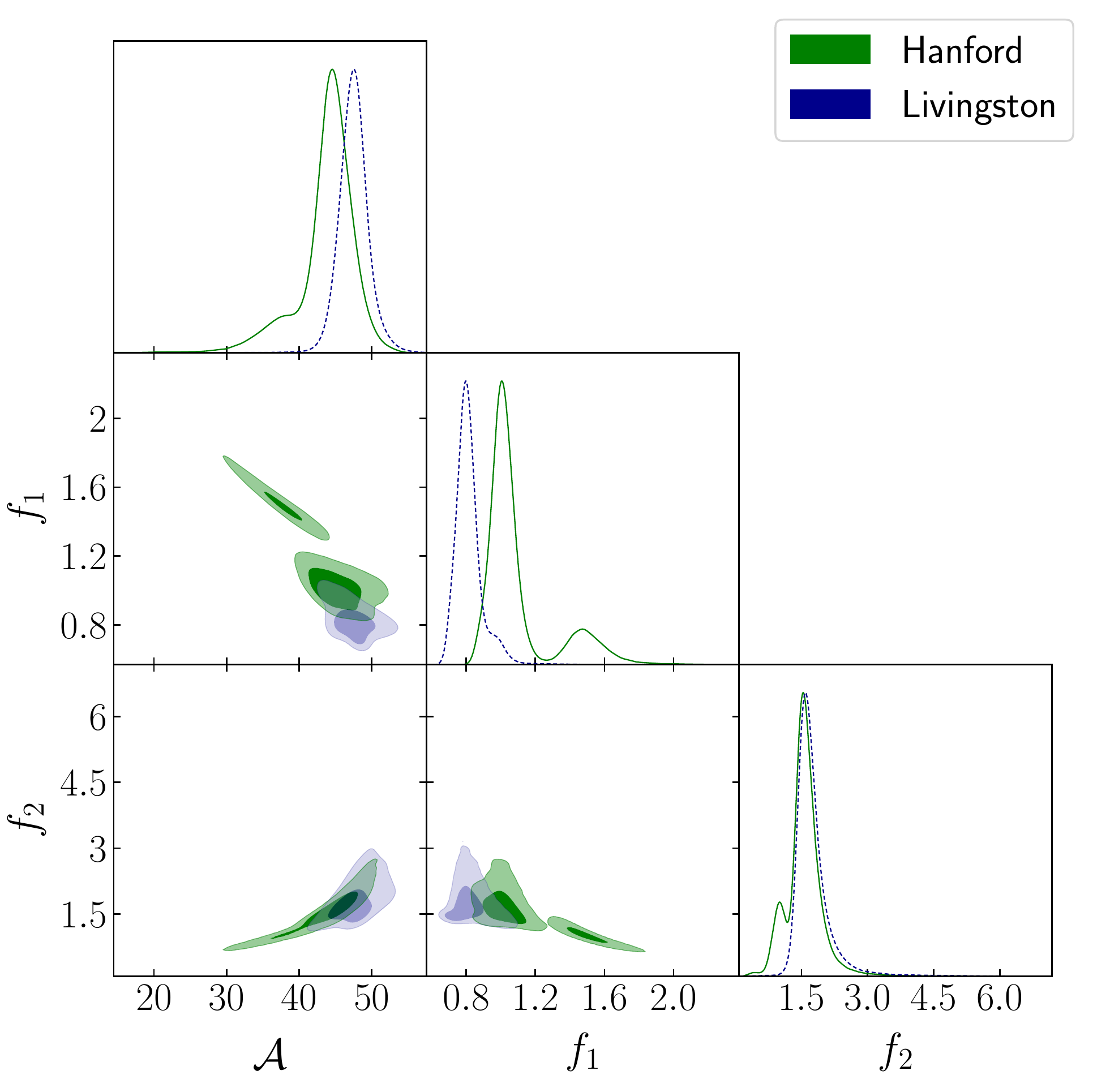}
\caption{\label{fig:triangle} One and two dimensional distributions in the noise fitting parameters; white level amplitude $\cal A$, and re-scaled knee frequencies $f_1$ and $f_2$. The distributions are marginalised in the excluded dimension. The distribution shows that the Hanford detector operated in two distinct sensitivity modes during the run characterised primarily by the binomial distribution in the $1/f$ component knee frequency. This splitting was correlated in time with the higher $f_1$ regime occurring in the last half of the run.}
\end{figure}

\begin{figure}[t]
\includegraphics[width = \linewidth]{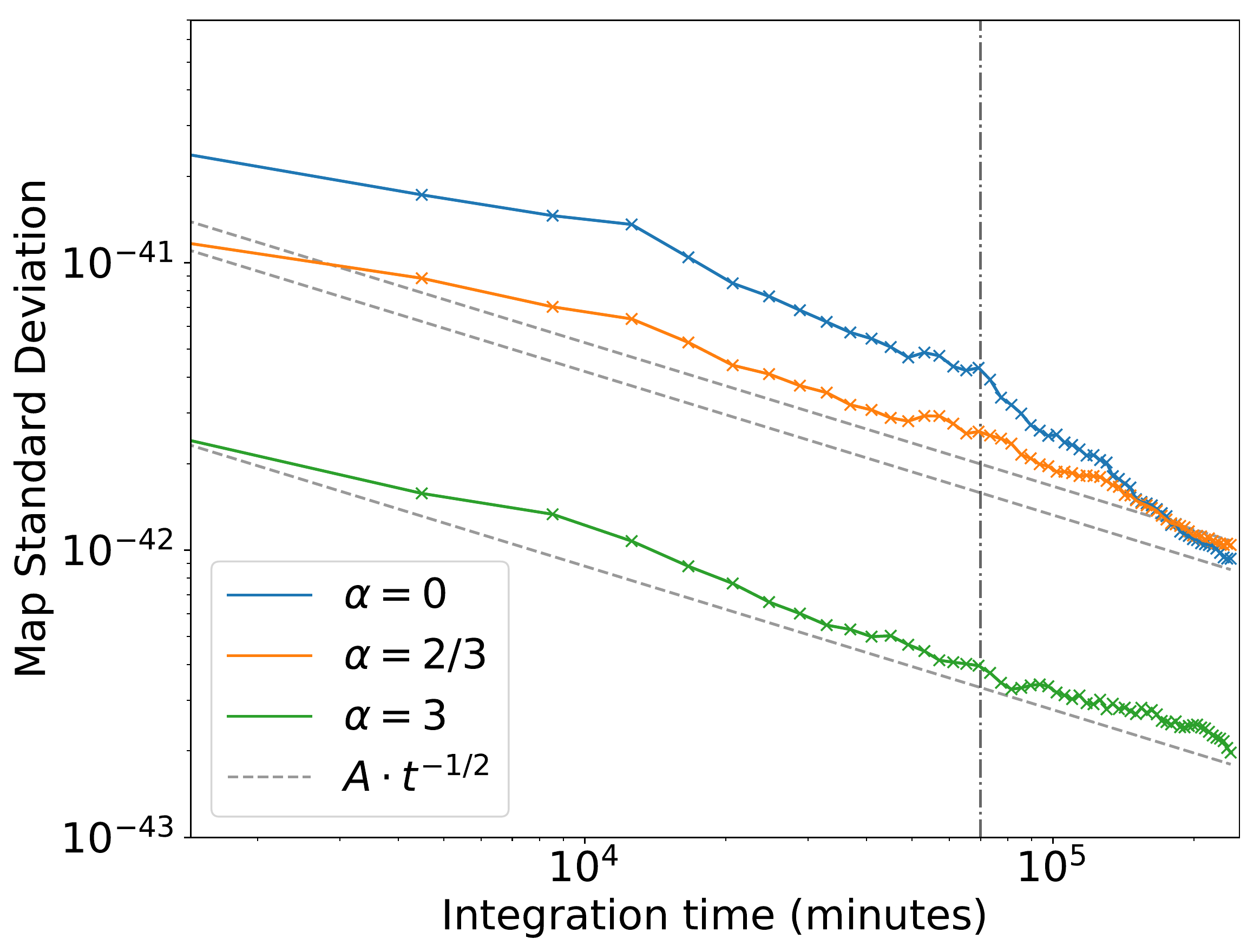}
\caption{\label{fig:stdev}Evolution of the standard deviation of the maximum-likelihood maps with integration time for $\alpha = 0,\,2/3,\,3$ cases. The constant slopes (short-dashed) are reference $t^{-1/2}$ power laws normalised at the end of the integration. The $\alpha=0$ and $2/3$ cases deviate from this constant noise approximation during the O2 run (right of vertical dashed line). This indicates that the sensitivity of the O2 run increased as a function of time. Given the spectral weighting of the two cases this also indicates that the improvement was in the low frequency $1/f$ component.}
\end{figure}

\section{Application to O1 and O2 data}\label{sec:data} 

\subsection{Noise Analysis}\label{sec:noise}

Although the O2 run data constitutes an overall improvement in sensitivity the characterisation of the noise is more complex than was found in the case of the O1 run, in part due to the increased duration of O2. A number of unidentified noise harmonics are present in the data and these also tend to wander in frequency in specific regions. These lines appear at different frequencies and severity in the two detectors; overall, the Hanford detector is most affected. Our noise fitting procedure attempts to minimise the impact of these effects since, although they would not bias the final signal estimate, they would lead to sub-optimal integration of the data.

Briefly, our noise fitting procedure is the following. We analyse all simultaneous 60 second segment time-streams from each detector separately. For each detector segment we calculate a windowed power spectrum and perform a three parameter curve fit to $P(f)$, with
\begin{equation}
P(f) = {\cal A}^2\left[\left(\frac{18 f_1}{0.1+f}\right)^4+\left(\frac{f}{50 f_2} \right)^2+(0.07)^2\right]\times10^{-44}\,.
\end{equation} 
Here, ${\cal A}$ is an overall normalisation factor which depends on the estimation process, and $f_1$ and $f_2$ are respectively the low and high knee frequencies of the model. The parameters are re-scaled so as to be approximately centred around 1. Fig.~\ref{fig:epsart} shows the fractional distribution of the individual detector power spectra over the 60 second segments and the corresponding range in model fits. The presence of wandering harmonics in the data is obvious when plotting the spectra in this way as they lead to broader ``forests'' between 60 and 80 Hz.

Fig.~\ref{fig:triangle} shows the one and two dimensional, marginalised distributions for the fitting parameters $\cal{A}$, $f_1$, and $f_2$ over all 60 second segments during the O2 run for both detectors. There is a significant correlation between $1/f$ knee frequency and the white noise amplitude as expected but we also see that the Hanford detector operated in two distinct modes with a shift to higher $1/f$ knee frequency during the second half of the run. This leads to an abrupt loss in sensitivity for the lower frequencies, but is limited to the final 20 days of observation. This is correlated to a drop in $\cal{A}$, as this parameter is estimated over the whole frequency range and normalisation must be preserved.

The evolution of the standard deviation in the maximum-likelihood maps as a function of integration time seen in Fig.~\ref{fig:stdev} is scaling slightly faster than the expected $\sim t^{-1/2}$ behaviour for the O2 data, indicating that there is an improvement in the $1/f$ noise component; this is reflected in the fact that the $\alpha=0$ and $\alpha=2/3$ runs are most sensitive to this effect since they are dominated by the lower frequencies.

\begin{figure*}[t]
\centering
\begin{minipage}{.32\textwidth}
  \centering
  \includegraphics[width=0.95\linewidth]{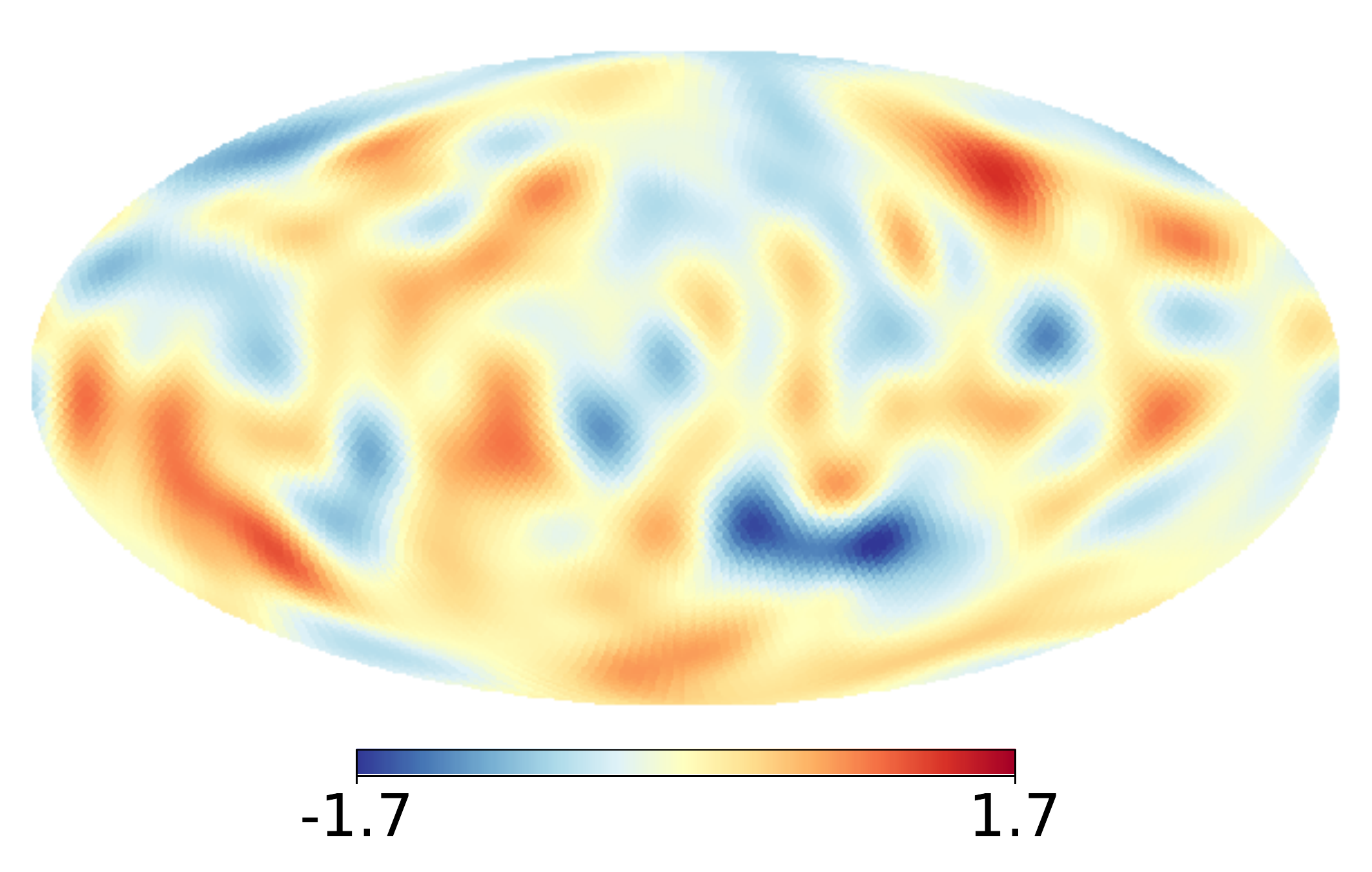}
  %\caption{A figure}
  %\label{fig:test1}
\end{minipage}%
\begin{minipage}{.32\textwidth}
  \centering
  \includegraphics[width=0.95\linewidth]{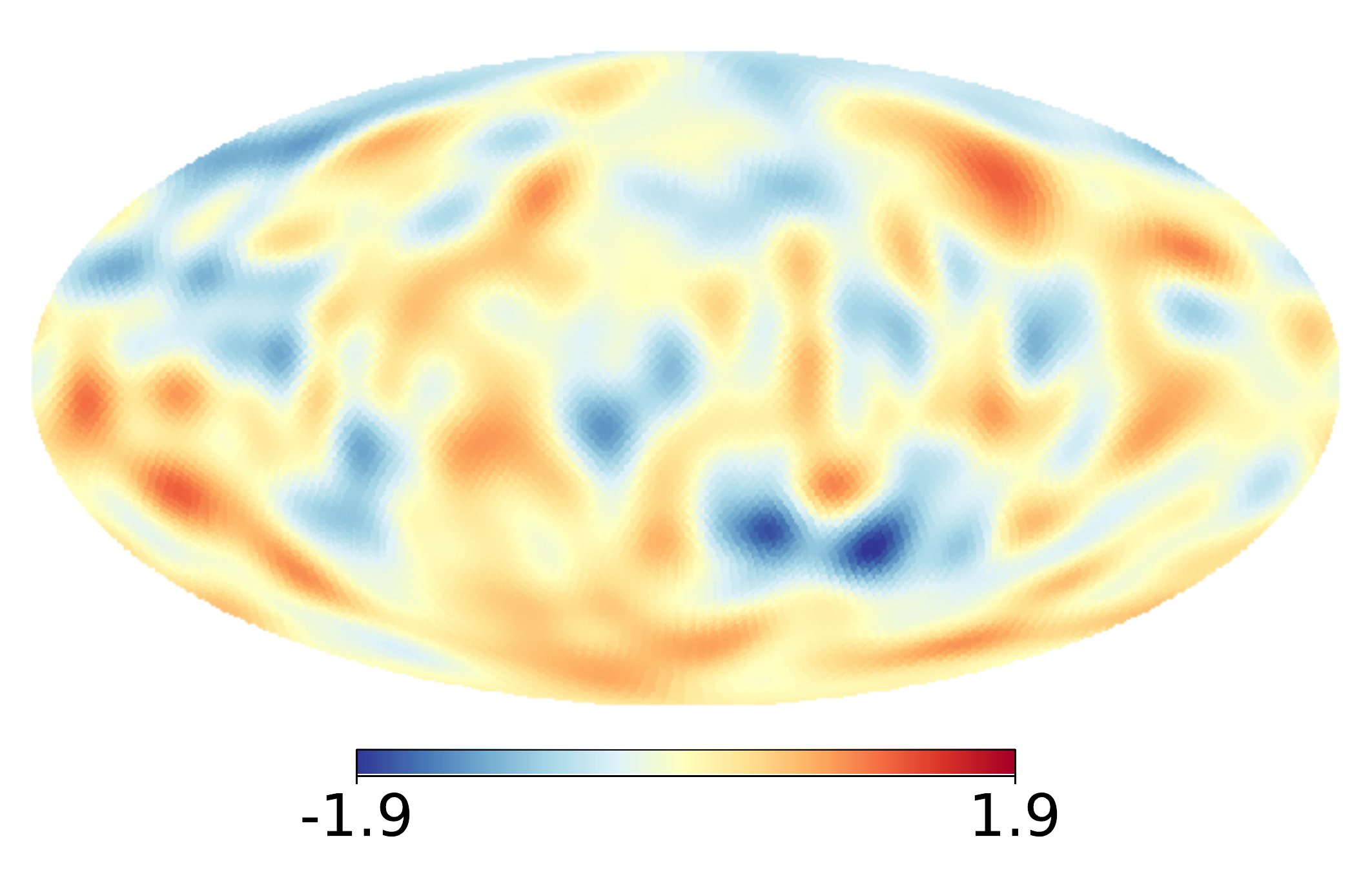}
  %\captionof{figure}{Another figure}
  %\label{fig:test2}
\end{minipage}
\begin{minipage}{.32\textwidth}
  \centering
  \includegraphics[width=0.95\linewidth]{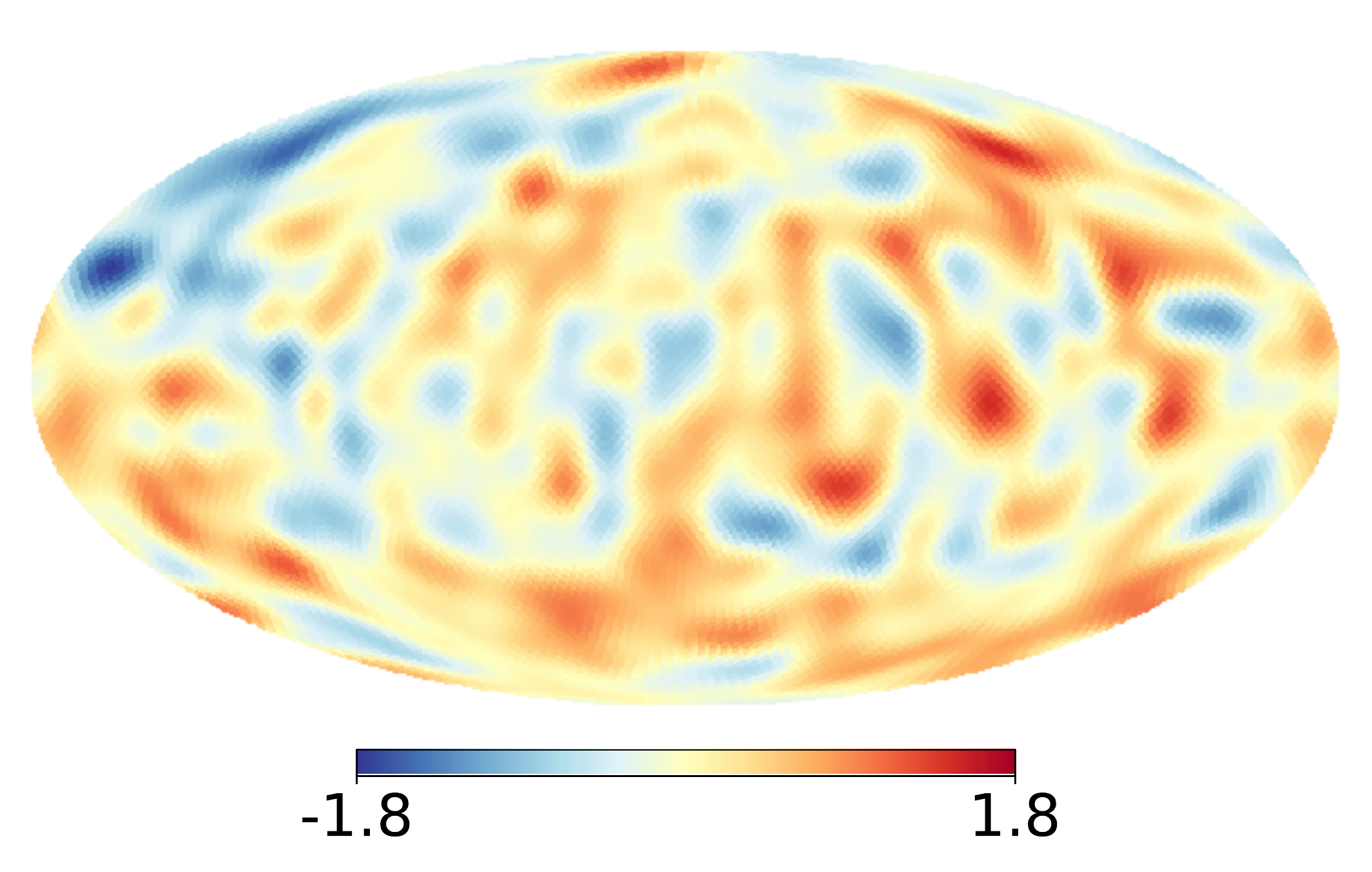}
  %\captionof{figure}{Another figure}
  %\label{fig:test2}
\end{minipage}
\begin{minipage}{.32\textwidth}
  \centering
  \includegraphics[width=0.95\linewidth]{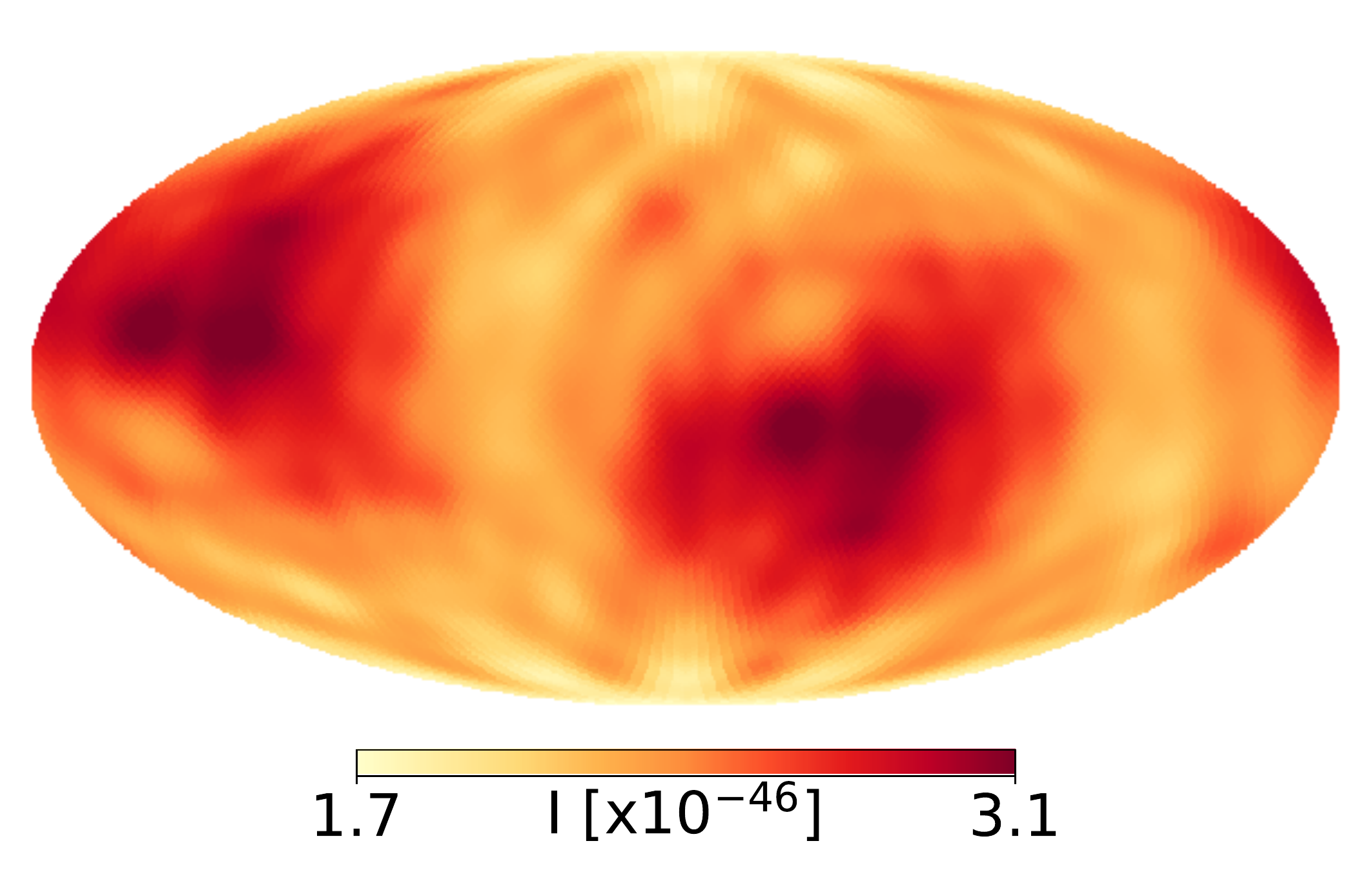}
  %\captionof{figure}{A figure}
  %\label{fig:test1}
\end{minipage}%
\begin{minipage}{.32\textwidth}
  \centering
  \includegraphics[width=0.95\linewidth]{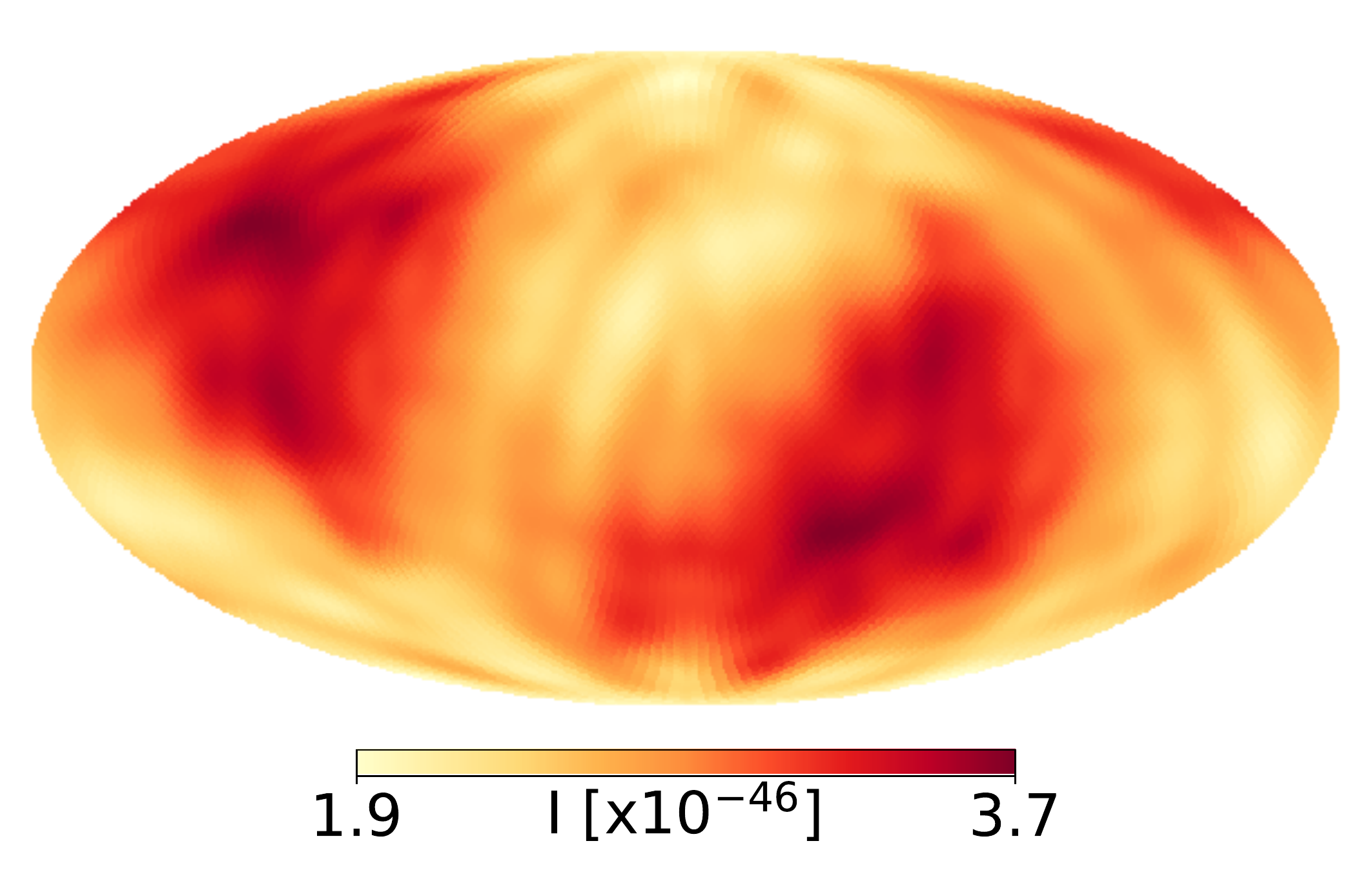}
  %\captionof{figure}{Another figure}
  %\label{fig:test2}
\end{minipage}
\begin{minipage}{.32\textwidth}
  \centering
  \includegraphics[width=0.95\linewidth]{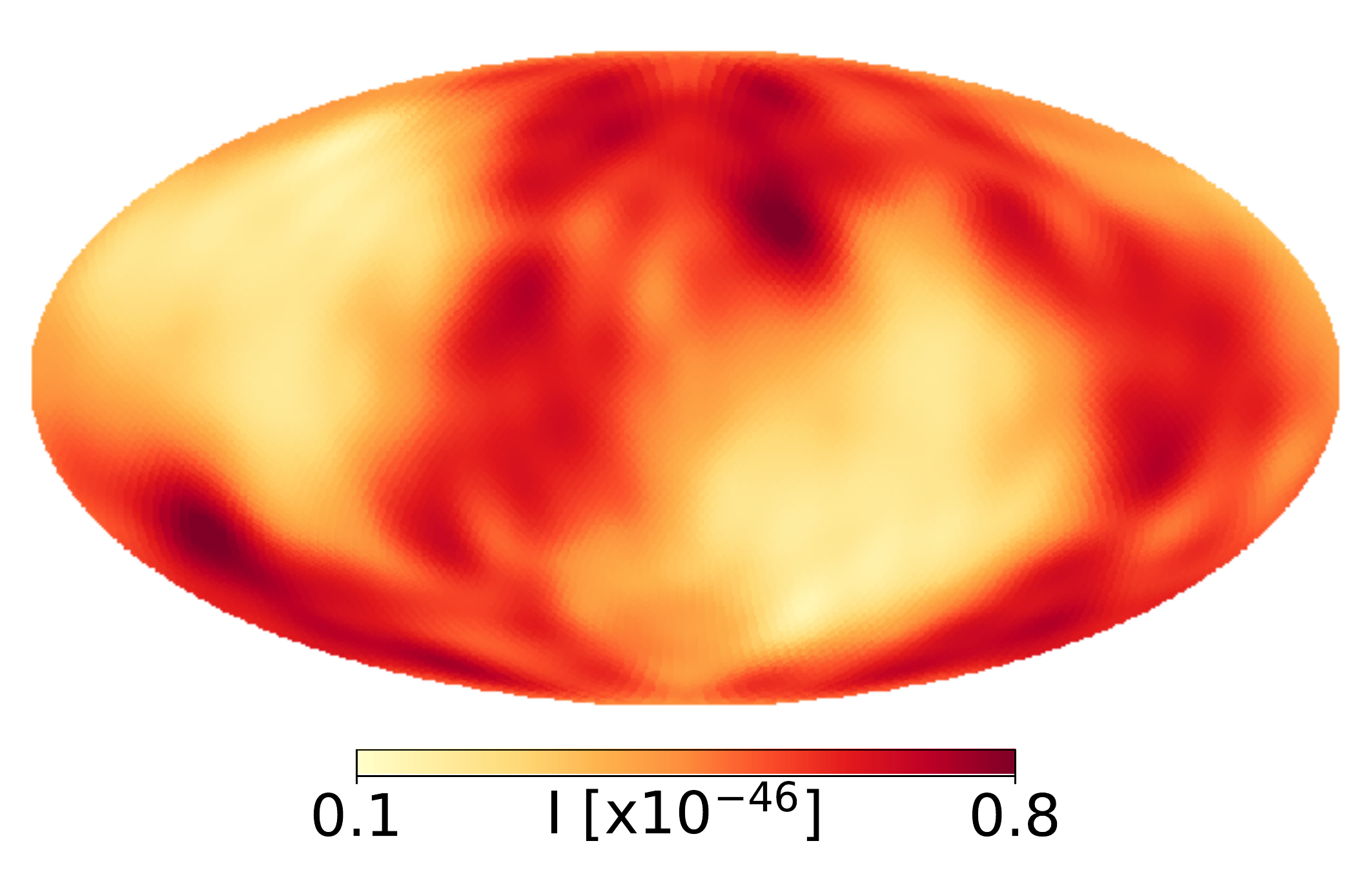}
  %\captionof{figure}{Another figure}
  %\label{fig:test2}
\end{minipage}
\caption{SNR (\emph{top}) and noise maps (\emph{bottom}) of the SGWB with spectral indices, from left to right, $\alpha = 0,\,2/3,\,3,$ respectively. All maps have been produced at a {\tt HEALPix} resolution $N_{\rm side}=8$.  For the purpose of visualisation we smooth the resulting maps with a 10 degree Gaussian FWHM beam. The $\alpha = 0$ and $2/3$ cases are at a reference frequency of $f_0 = 50$ Hz. The $\alpha = 3$ case has $f_0 = 100$ Hz. The noise maps are in units of Hz$^{-1}$. The SNR maps show, qualitatively, that the features in the maximum-likelihood solutions are consistent with the noise covariance.
}
\label{fig:maps}
\end{figure*}

\subsection{Model-dependent, broadband limits}\label{broad}

We first present the results from a broadband integration using three different spectral weightings, $\alpha = 0,\,2/3,\,3$. In this case we integrate the entire range of frequencies into a single map for each choice of $\alpha$. As such, the map $z_p$ and operator $M_{pp'}$, in (\ref{method}) and (\ref{method2}) respectively, are accumulated over $f=[30,\,480]$ Hz. The results are shown in Fig.~\ref{fig:maps}. Our convention is to show signal-to-noise (SNR) and noise maps since our results are only constraints on any signal level. The SNR maps are obtained by normalising the maximum-likelihood maps obtained at the end of the integration by the pixel covariance matrix given by ${\cal N}_{pp'} \equiv M_{pp'}^{-1}$,
\begin{equation}
    S_{p} = \sum_{p'}\, {\cal N}_{pp'}^{-1/2}\, I_{p'}\,,
\end{equation}
where ${\cal N}_{pp'}^{-1/2}$ is the inverse of the Hermitian square root of the covariance matrix. This results in a map in units of the expected standard deviation that can be inspected visually for outliers. The off-diagonal pixel to pixel correlations are significant however and therefore an interpretation just based on the SNR maps is only qualitative. The maps are smoothed to a common resolution using a Gaussian filter of 10 degrees FWHM to reduce pixel noise. 

The noise maps shown in Fig.~\ref{fig:maps} are obtained by plotting the diagonal elements of ${\cal N}$. These show how the pixel variance changes across the sky given the full time and frequency integration. The maps also do not carry information about the significant off-diagonal correlations. The variance structure in the maps changes significantly as a function of the spectral index used for the signal weighting in the projectors. To quantify the consistency of the final maps with the expected noise covariances we calculate $\chi^2$ statistics using the full covariance matrices and report probability to exceed statistics ${\cal P}_{\rm ex}$ based on these in Table~\ref{tab:results}.
\begin{table}[t]
\begin{center}
\caption{Constraints on the isotropic background amplitude for different target spectral indices $\alpha$. The integration includes frequencies between 30 and 480 Hz.}
\label{tab:results}
\begin{tabular}{ c | c | c | l | l | l }
\hline
\rule{0pt}{2.6ex}\rule[-1.2ex]{0pt}{0pt}&$\alpha$ & $f_0$ [Hz]  & $\Omega_{\rm GW} (f_0)$& 95\%  {\sl u.l.} & $\mathcal{P}_{\rm ex}$   \\

\hline
\hline
%&&&\\
\rule{0pt}{2.6ex}&  0 & 50 &  $(-0.3 \pm 1.4) \times 10^{-7}$ & $2.6\times 10^{-7}$ & 77\%  \\
O1&2/3 & 50 & $(-7.9 \pm 8.1)\times 10^{-8}$ & $8.2\times 10^{-8}$ & 86\% \\
\rule[-1.2ex]{0pt}{0pt}&3 & 100 &$(2.1 \pm 2.0) \times 10^{-7}$ &  $5.9 \times 10^{-7}$ & 41\% \\ 
\hline
\hline
%&&&\\
\rule{0pt}{2.6ex} O1&  0 & 50 &  $(-2.5 \pm 3.9) \times 10^{-8}$ & $5.5\times 10^{-8}$ & 19\%  \\
 + &2/3 & 50 & $(-1.6 \pm 3.4)\times 10^{-8}$ & $5.2\times 10^{-8}$ & 70\% \\
\rule[-1.2ex]{0pt}{0pt} O2&3 & 100 &$(1.5 \pm 0.9) \times 10^{-7}$ &  $3.2 \times 10^{-7}$ & 28\% \\ 
\hline
\end{tabular}
\end{center}
\end{table}

\subsubsection{SGWB Monopole limits}

Our estimated intensity maps $I({f_0,\hat p})$ yield an estimate of $\Omega_{\rm{GW}}$ at the reference frequency $f_0$ subject to the broad band integration using an assumed spectral index $\alpha$. $\Omega_{\rm{GW}}$ itself is related to the energy density $\rho_{\rm GW}$ of the gravitational-waves per logarithmic frequency interval as
\begin{equation}
\Omega_{\rm GW}(f) = \frac{1}{\rho_c}\frac{d\rho_{\rm GW}(f)}{d\ln f}\,,
\end{equation}
where $\rho_c$ is the critical energy density of the Universe. 

\begin{figure}[t]
\includegraphics[width = \linewidth]{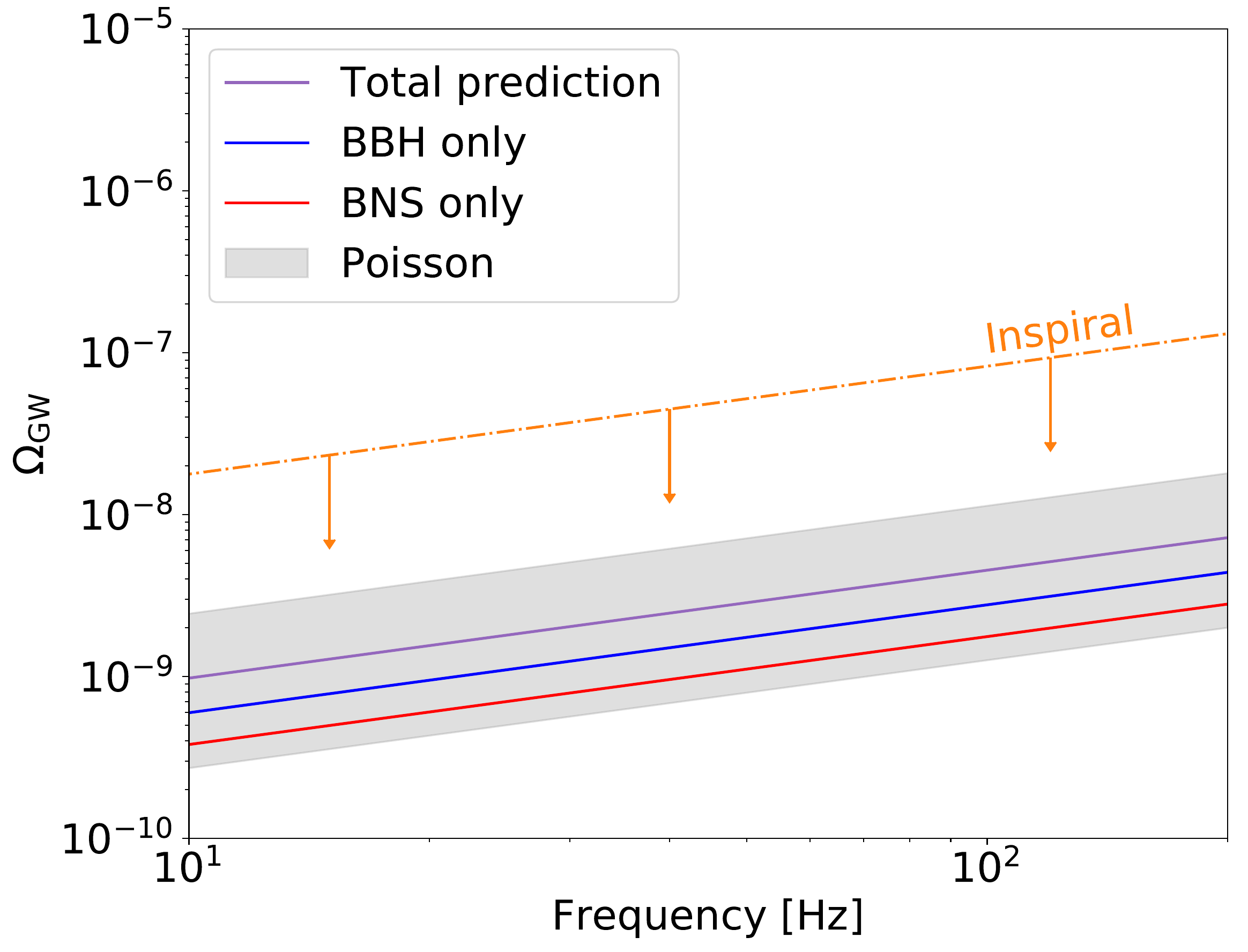}
\caption{Our 95\% upper limit (dashed) for the broadband integration assuming the spectral index $\alpha=2/3$ compared to predictions for the stochastic background due to black hole-black hole (BBH) and black hole-neutron star (BNS) binary system inspirals \cite{LIGOimpli2018}. The shaded area is an estimate of the uncertainty in the model prediction and the plot is cutoff at 200 Hz as the background is expected to fall off sharply beyond that frequency.
}
\label{fig:ins_pred}
\end{figure}

The monopole, $I(f_0)$, of the maps is calculated using a noise weighted average of the pixels in $I_p$. Note that these are {\sl not} specific intensity maps and therefore background quantities are obtained by averages over the maps. This can be scaled to yield an estimate of $\Omega_{\rm{GW}}$ using~\cite{Allen1996}
\begin{equation}
 \Omega_{\rm GW}(f_0)  =  \frac{4\pi^2}{3H_0^2}  f_0^3  I(f_0)\,,
\end{equation}
where $H_0 = 70$ km s$^{-1}$ Mpc$^{-1}$ is the Hubble rate today. Note that the intensity monopole $I(f_0)$ also includes a normalisation by a factor of $5/8\pi$ due to the normalisation of the detector response functions used in projection operator $A_{pf}$ (see for example \cite{Allen1999}). The result can be scaled to arbitrary frequencies using the same spectral index assumed in the estimation using
\begin{equation}
    \Omega_{\rm GW}(f) = \Omega_{\rm GW}(f_0)\left(\frac{f}{f_0}\right)^\alpha\,.
\end{equation}
Limits on $\Omega_{\rm{GW}}$ for the three choices of spectral indices at specific reference frequencies are shown in Table~\ref{tab:results}. None of the choices show any significant excess over the expected noise\footnote{Note that the limits quoted for O1 are slightly different than those presented in~\cite{Renzini2019}; this is due to correction in the effective weighting being used which was incorrectly eliminating high frequencies.}.
The limits are many orders of magnitude above the amplitude expected of a primordial background produced during inflation \cite{Grishchuk1993} with scale invariant  $\Omega_{\rm{GW}} \sim {\cal O}(10^{-14})$. The detection of a stochastic background of astrophysical origin may not be that far off however. In particular, the $\alpha=2/3$ spectrum choice is motivated by forecasts of the stochastic signal of binary system inspirals, either binary black hole or black hole-neutron star systems.  Fig.~\ref{fig:ins_pred} shows our upper limit compared to the predictions of \cite{LIGOimpli2018}. The highest estimates of the stochastic signal are only one order of magnitude below the current upper limit. The limits on $\Omega_{\rm{GW}}$ will scale linearly with individual detector strain sensitivities and therefore the background signal may be within reach of the next planned LIGO update \cite{Christensen2018} with similar integration times.

\subsubsection{SGWB Anisotropies}
Despite the lack of evidence of any signal it is worth describing how to obtain limits on any anisotropy about the mean background. Anisotropies are most often parametrised in terms of the coefficients $a_{\ell m}$ of the spherical harmonic expansion of the sky map
\begin{equation}
    a^{\,}_{\ell m}(f_0) = \int d\Omega_{\hat p}\, I(f_0,\hat p)\,Y^\star_{\ell m}(\hat p)\,,
\end{equation}
where the $Y_{\ell m}(\hat p)$ are the spherical harmonic basis functions. In our case, since our maximum-likelihood solution is obtained in the pixel domain, we need to expand both the sky maps and the noise covariances in spherical harmonics in order to obtain optimal estimates of the $a_{\ell m}$. Since the map and $a_{\ell m}$'s are related by a linear projection it is simple to show that the maximum-likelihood estimate of the vector of coefficients $a$ given a map $I_p$ with covariance ${\cal N}$ is given by (using compact notation)
\begin{equation}\label{eq:rotate}
    a = \left({\cal Y}^\dagger {\cal N}^{-1} {\cal Y}\right)^{-1}\, {\cal Y}^\dagger{\cal N}^{-1} I\,,
\end{equation}
where the linear operator ${\cal Y}_{\ell m, p}\equiv Y_{\ell m}(\hat p)$ and similar for its adjoint.

\begin{figure}[t]
\includegraphics[width = \linewidth]{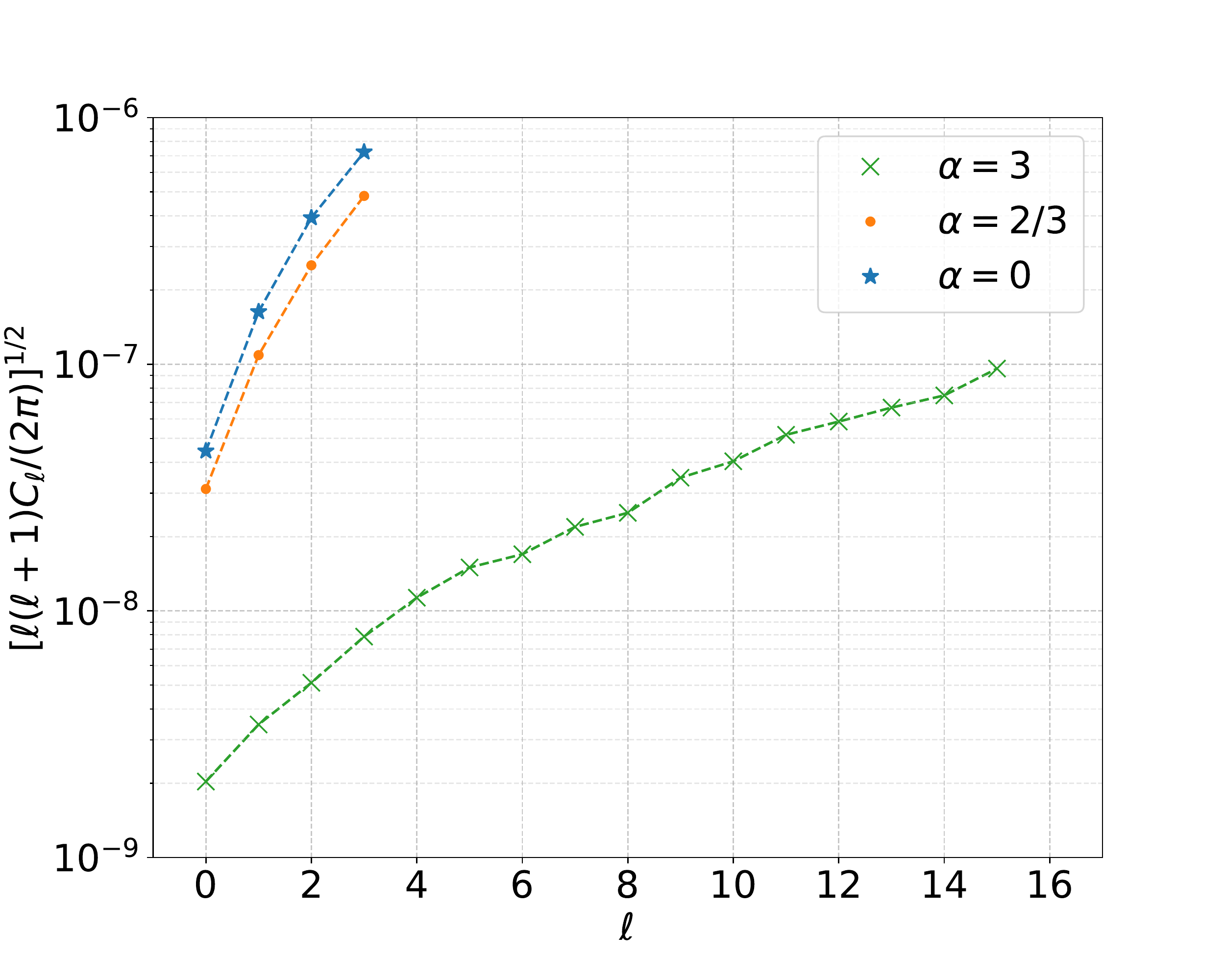}
\caption{95\%  upper limits on the signal angular power spectrum $C_\ell$'s of the SGWB for the the three assumed spectral indices. All values are scaled to a reference frequency $f_0 = 50$ Hz. Depending on the spectral index, the estimate converges up to different maximum multipoles $\ell$ with the $\alpha=3$ case giving the highest resolution since it integrates low and high frequency with equal weights. We show $\ell(\ell+1) C_{\ell}/2\pi$ since this is a measure of equal variance per logarithmic interval in $\ell$. A scale invariant signal in this measure would be flat.}
\label{fig:cells}
\end{figure}

If the signal is expected to be statistically isotropic (i.e. no preferred direction) then the coefficients can be compressed to an angular power spectrum using an unbiased estimator
\begin{equation}
    C_\ell(f_0) = \frac{1}{2\ell + 1}\sum_m\, |a_{\ell m}(f_0)|^2 - \tilde {\cal N}_\ell\,,
\end{equation}
where the isotropised noise bias $\tilde {\cal N}_\ell$ can be computed from an expansion of ${\cal N}$,
\begin{equation}\label{eq:rotate2}
    \tilde {\cal N}_\ell  = \frac{1}{2\ell + 1}\sum_m\,|\sum_{p p'} {\cal Y}^{\,}_{\ell m, p}{\cal N}^{-1}_{pp'}{\cal Y}^\star_{p',\ell m}|^2\,.
\end{equation}
Constraints on the $C_\ell$'s, scaled to $\Omega_{\rm GW}$ for the three separate spectral indices, are shown in Fig.~\ref{fig:cells}. The choice of maximum multipole $\ell$ in (\ref{eq:rotate}) and (\ref{eq:rotate2}) is important; a value that is too low will not exploit the available resolution in the maps, a value that is too high will lead to ill-conditioning which will be manifested as aliasing in the final spectra. We increase the maximum $\ell$ until the spectra have converged at low multipoles, typically $\ell\sim 16$ and truncate at an $\ell$ above which the spectra are clearly aliased. The resulting truncation depends on the choice of spectral index with $\alpha=3$ containing the highest resolution. This is in agreement with the structure shown in Fig.~\ref{fig:maps} and is a consequence of the spectral weighting being flat in intensity. 

\subsection{Model-independent spectral limits}\label{sec:segm10}
An alternative approach to signal estimation is to reduce, to a minimum, the assumptions made about the spectral dependence of the stochastic background. This approach is motivated if we suspect that the background may be populated by distinct sources with different spectral shapes - this would lead to breaks in the power-law for the background or even more complicated features such as well-defined peaks \cite{Hogan1986,Caprini2019}. In fact, a model-independent spectral estimation as we propose here could be part of a method aimed at separating the contributions of different sources based on their spectral dependence. Combined with the angular separation provided by the mapping itself, this constitutes a prototype separation method for future observations where more than one family of galactic and extra-galactic sources will be visible in the target frequency range.  

To test this method we run our algorithm on ten separate frequency bins, accumulating the projection operators only over frequencies assigned to each bin. We assume a spectral dependence within each bin that is scale-invariant with respect to the dimensionless background density i.e. $E(f)\sim 1/f^3$. This is a form of ``least informative" prior on the spectral dependence but other choices can be made in order for the compression to be optimal with respect to a specific spectral shape \cite{Bond2000}.

\begin{figure}[t]
\includegraphics[width = \linewidth]{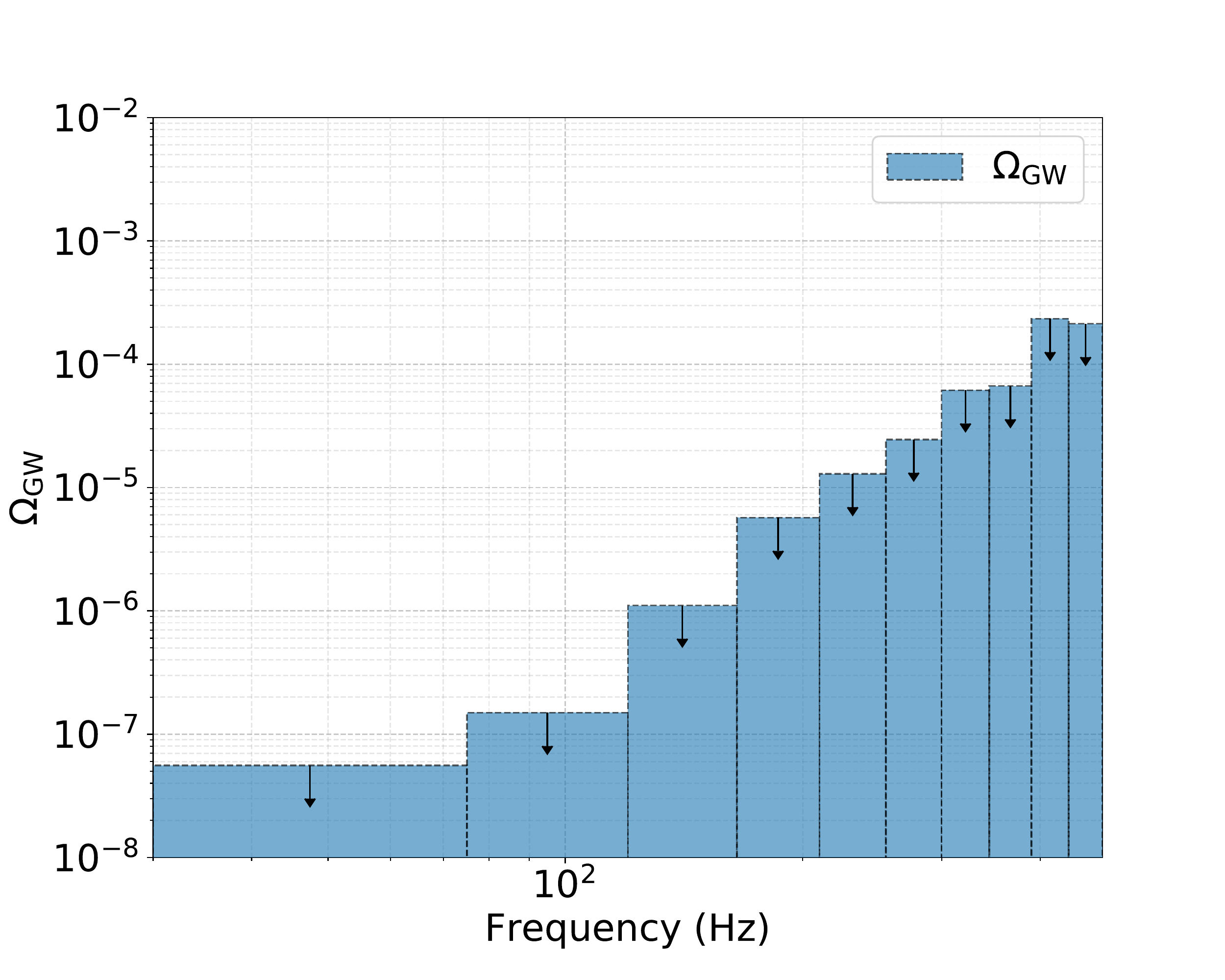}
\caption{95\% upper limits for the SGWB monopole for each of the maps obtained in the ten separate frequency bins used in the model-independent spectral analysis. The method assumes a scale invariant shape for the signal in each spectral bin. The spectral shape is consistent with a noise dominated estimate.}
\label{fig:10segs}
\end{figure}

\begin{figure*}[t]
\centering
\begin{minipage}{.32\textwidth}
  \centering
  \includegraphics[width=0.95\linewidth]{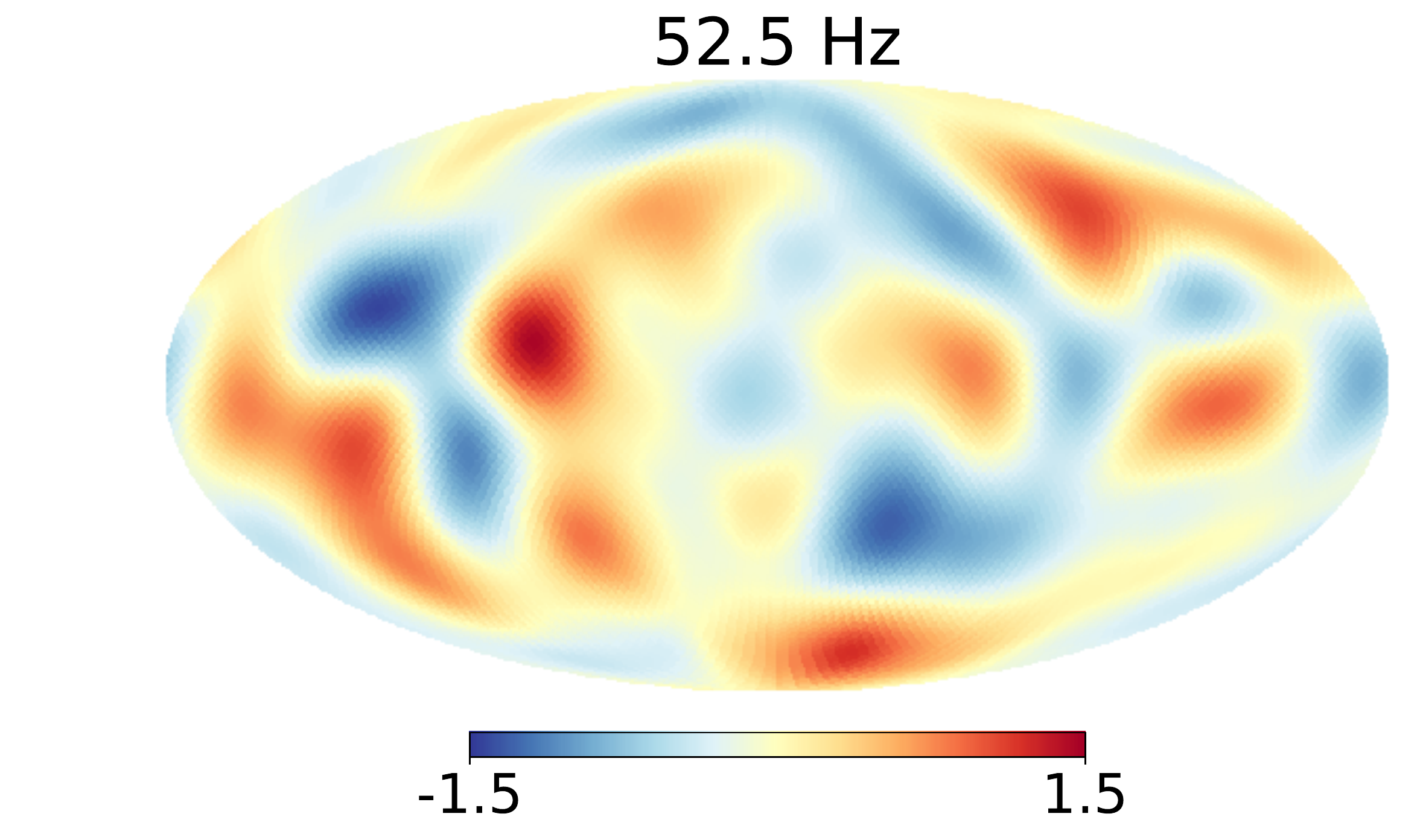}
  %\caption{A figure}
  %\label{fig:test1}
\end{minipage}%
\begin{minipage}{.32\textwidth}
  \centering
  \includegraphics[width=0.95\linewidth]{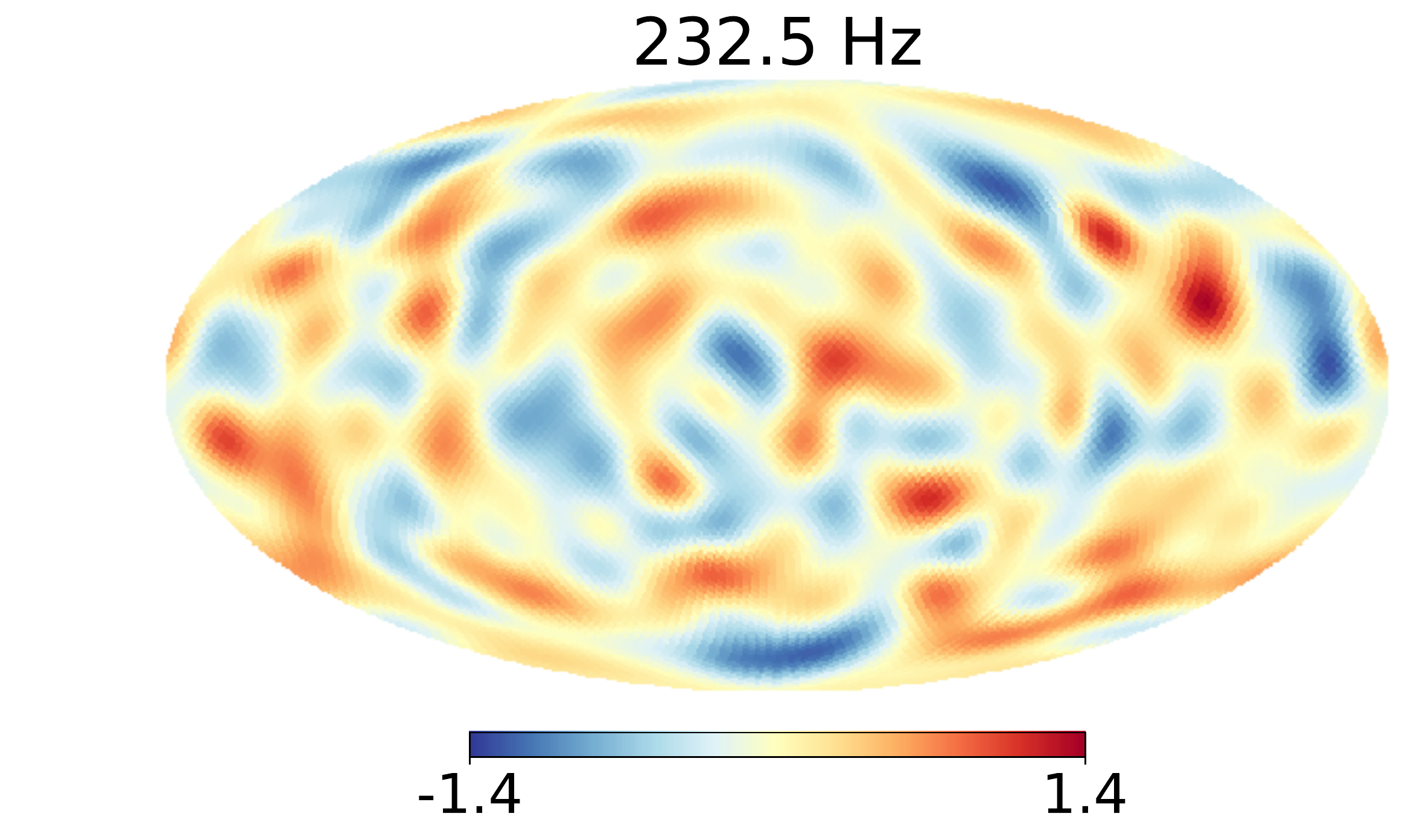}
  %\captionof{figure}{Another figure}
  %\label{fig:test2}
\end{minipage}
\begin{minipage}{.32\textwidth}
  \centering
  \includegraphics[width=0.95\linewidth]{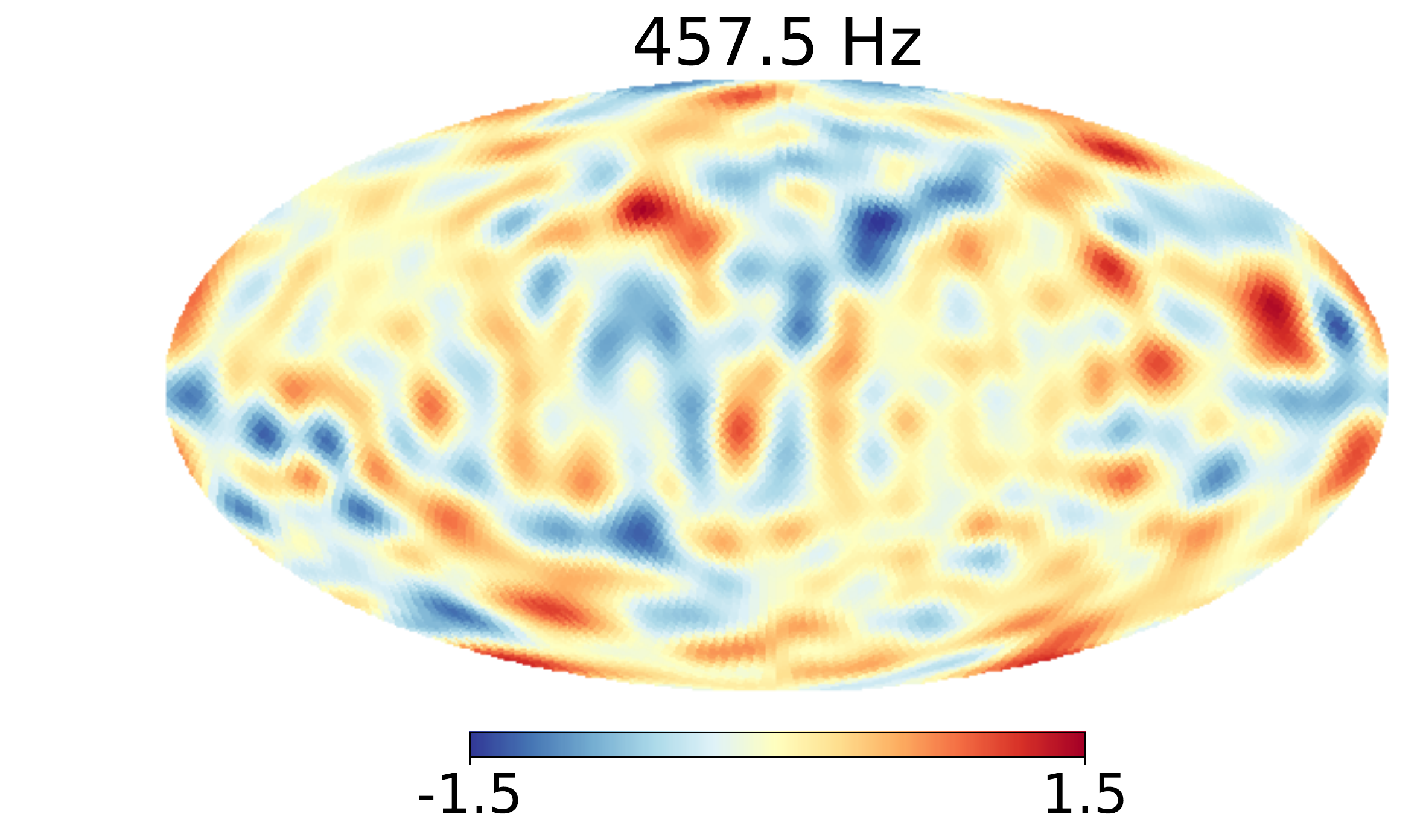}
  %\captionof{figure}{Another figure}
  %\label{fig:test2}
\end{minipage}
\begin{minipage}{.32\textwidth}
  \centering
  \includegraphics[width=0.95\linewidth]{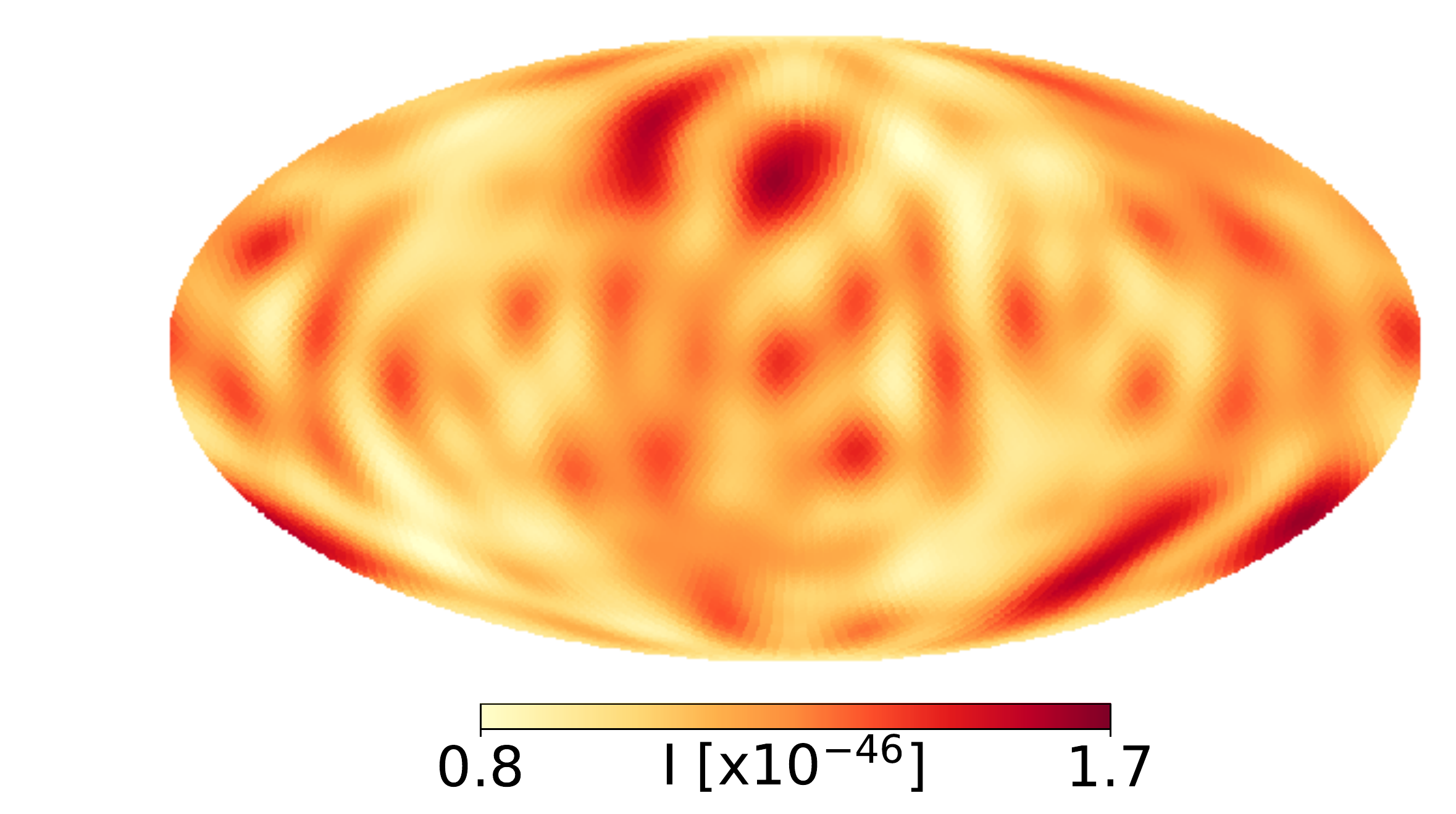}
  %\captionof{figure}{A figure}
  %\label{fig:test1}
\end{minipage}%
\begin{minipage}{.32\textwidth}
  \centering
  \includegraphics[width=0.95\linewidth]{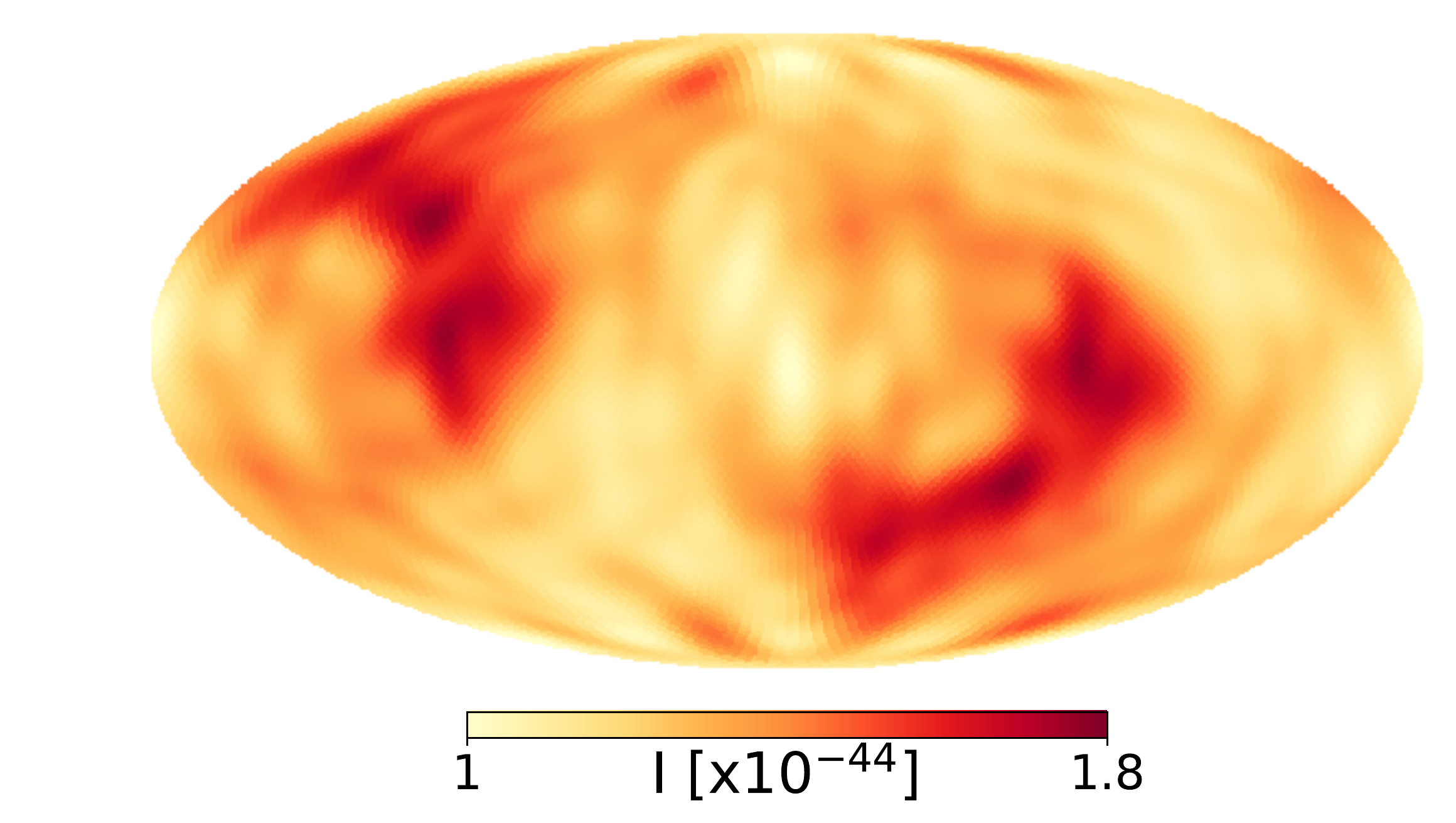}
  %\captionof{figure}{Another figure}
  %\label{fig:test2}
\end{minipage}
\begin{minipage}{.32\textwidth}
  \centering
  \includegraphics[width=0.95\linewidth]{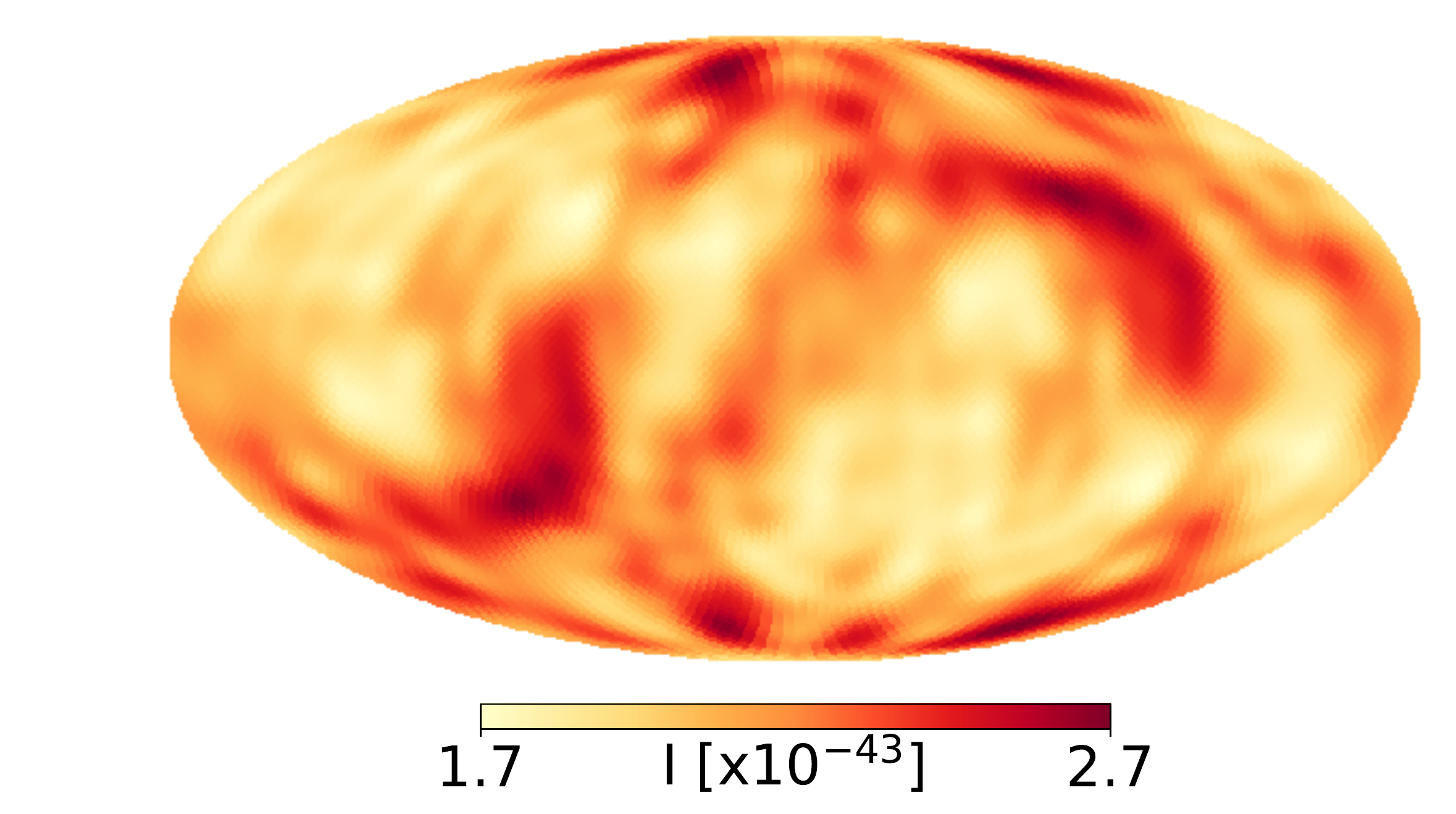}
  %\captionof{figure}{Another figure}
  %\label{fig:test2}
\end{minipage}
\caption{SNR (\emph{top}) and noise maps (\emph{bottom}) for three of the 45 Hz--wide frequency bins used in the model-independent spectral decomposition, centred, from left to right, around 52.5 Hz, 232.5 Hz, 457.5 Hz, respectively. The structure in the noise maps depends significantly on the frequency, as expected. This also drives the apparent structure in the SNR maps which is purely noise dominated.}
\label{fig:maps2}
\end{figure*}

The bins are defined as ten equally spaced intervals of width 45 Hz between 30 and 480 Hz. Ten maps are produced in this exercise, one for each frequency bin, with associated noise covariance matrices. As described in Section~\ref{broad} we can estimate both a $\Omega_{\rm GW}$ and its anisotropies from each map. The results for the 95\% upper limit in the monopole of $\Omega_{\rm GW}$ are shown in Fig.~\ref{fig:10segs}. The constraints are higher than the signal in the equivalent single, broadband estimate, since the information is split into ten separate estimates but it is useful to note that the spectral dependence is consistent with a noise dominated estimate with increasing power as a function of frequency.
Fig.~\ref{fig:maps2} shows the SNR and noise maps for three of the frequency bins. Given the noise level in this analysis we do not extract upper limits on any measures of the anisotropy but it is useful to note how the mode structure on the sky changes as a function of frequency. This effect, evident in the top half of Fig.~\ref{fig:maps2}, becomes very apparent in this kind of spectral analysis and can be understood through the frequency-to-sky mode coupling encoded in the projection operator $A_{pf}^\tau$ \cite{Renzini2018}. The bottom half reveals the sky modulation of the noise as a function of frequency, which appears smoothed in the noise maps in Fig.~\ref{fig:maps}. 
\section{\label{sec:disc}Discussion}

The upper limits for $\Omega_{\rm GW}(f_0)$ obtained by integrating O1 and O2 data runs presented in Table~\ref{tab:results} are the most constraining to date. They are in agreement with independent results presented by the LIGO and Virgo collaboration~\cite{TheLIGOScientificCollaboration2019mono}, taking into consideration longer integration times used here, as in our analysis we discard fewer time segments overall. This difference is probably due to the independent quality control pipelines but also due to the shorter segment length we adopt for our analysis. This last point means that, in principle, the stationarity conditions are more easily satisfied in our analysis. Comparing results between runs it is apparent that the 1/$f$ component of the noise is significantly better in the O2 run given the improvements in the $\alpha=0$ case which is the most sensitive to lower frequencies.  Despite the fluctuation in noise regimes during the O2 run, the duration and overall sensitivity improvements mean that it dominates the signal-to-noise integration of the combined run.

The SNR scaling in the maps in Fig.~\ref{fig:maps} and the time evolution of the standard deviation of the sky strain intensity plotted in Fig.~\ref{fig:stdev} only show the noise distribution and the diagonal pixel-pixel correlation on the sky. As such it is not meaningful to compare them with those presented in~\cite{TheLIGOScientificCollaboration2019} from the LIGO and Virgo collaboration, as these distributions depend substantially on the different methods used in the integration of the data. A quantitative comparison of the output of the two independent pipelines will become more important as we approach detection. The small scales present in the maps in Figs.~\ref{fig:maps} and~\ref{fig:maps2} are then simply the scales at which the noise fluctuates for a given frequency weighting and pixel size, and are not to be mistaken with the angular resolution of the detectors.
%The noise-dominated regime is also apparent from %the power in the $C_{\ell}$s, as the upper limits %shown in Fig.~\ref{fig:cells} increase as a %function of $\ell$ on the sky, which underlines %high correlation between power and scale of the %noise.

Despite the results being presented here being upper limits it is useful to note two important contributions of our work. The first is the consistent evaluation of both the monopole of the background {\sl and} its anisotropies in generalised sky coordinates. This means our method is ideally suited for the future regime where any number of detectors will be networked to provide a number of cross-correlation time-streams. This method is applicable to both ground or space-based detectors, or indeed, a mixture of the two as long as frequency bands can be overlapped with sufficient synchronous precision.

Secondly, we have shown how these methods can be employed to obtain a spectral breakdown of the signal. Once again, this will be very useful once we enter the high signal-to-noise regime when multiple backgrounds may be detected simultaneously. Our method is a basis for the {\sl combined} angular/frequency based separation of multiple backgrounds.

Alternatively, our method can be used to explore the efficiency of future network configurations in mapping the background. Forecasting the effective mode coverage of a network of gravitational wave interferometers is complicated by the non-compact beams of each detector. It is a much harder than, for example, radio interferometry where the primary beam is compact on the sky and the flat-sky approximation can be used along with assumptions of finite resolution in $uv$-space.

The interest in gravitational wave astronomy and cosmology is driving the development of detectors with increased sensitivity and the construction of new facilities around the world. This heralds an exciting future for gravitational wave science and the mapping methods developed in this line of work will be pivotal to the analysis of future data sets.

\begin{acknowledgments}
This research was supported by Science and Technology Facilities Council consolidated grants ST/L00044X/1 and ST/P000762/1. AIR acknowledges support of an Imperial College Schr\"{o}dinger Fellowship.
\end{acknowledgments}

\appendix

%\section{Appendixes}

\bibliography{refs.bib}% Produces the bibliography via BibTeX.
\bibliographystyle{apsrev}

\end{document}